# Scalar Gaussian Wiretap Channel with Peak Amplitude Constraint: Numerical Computation of the Optimal Input Distribution

Luca Barletta*, Alex Dytso**
* Politecnico di Milano, Milano, 20133, Italy, Email: luca.barletta@polimi.it
** New Jersey Institute of Technology, Newark, NJ 07102, USA, Email: alex.dytso@njit.edu

**Abstract—This paper studies a scalar Gaussian wiretap channel where instead of an average input power constraint, we consider a peak amplitude constraint on the input. The goal is to obtain insights into the secrecy-capacity and the structure of the secrecy-capacity-achieving distribution. Capitalizing on the recent theoretical progress on the structure of the secrecy-capacity-achieving distribution, this paper develops a numerical procedure, based on the gradient ascent algorithm and a version of the Blahut-Arimoto algorithm, for computing the secrecy-capacity and the secrecy-capacity-achieving input and output distributions.**

## I. Introduction

Consider the scalar Gaussian wiretap channel with outputs

$$Y_1 = X + N_1, \tag{1}$$

$$Y_2 = X + N_2, \tag{2}$$

where $N_1 \sim \mathcal{N}(0, \sigma_1^2)$ and $N_2 \sim \mathcal{N}(0, \sigma_2^2)$, and with $(X, N_1, N_2)$ independent of each other. The output $Y_1$ is observed by the legitimate receiver whereas the output $Y_2$ is observed by the malicious receiver. The block diagram for the Gaussian wiretap channel is shown in Fig. 1.

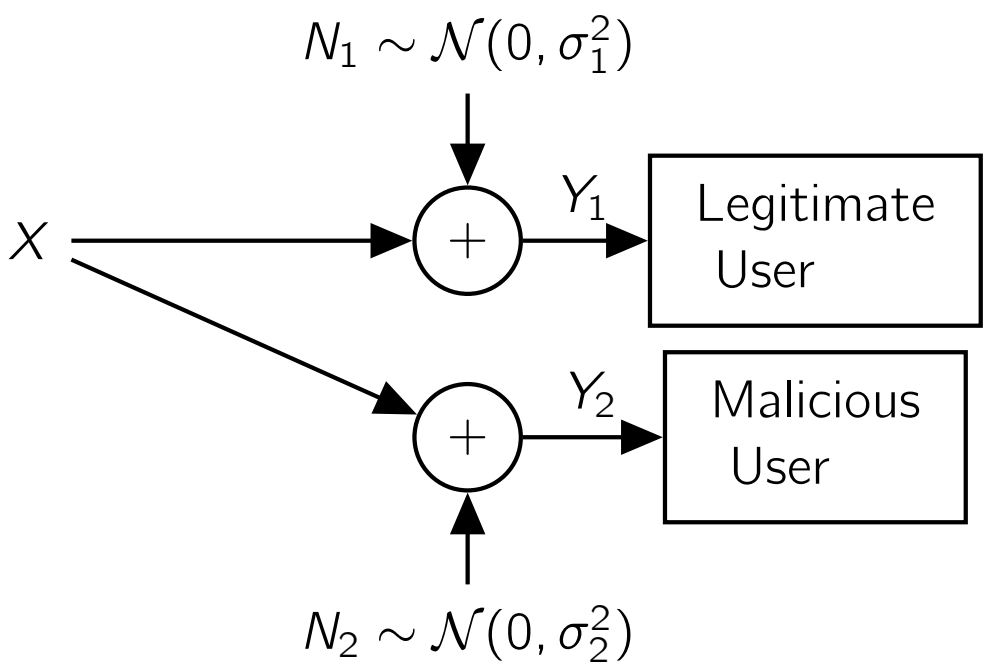

Fig. 1: The Gaussian wiretap channel.

In this work, we assume that the input $X$ is limited by a peak-power constraint or, equivalently, by a peak amplitude constraint given by $|X| \le \mathsf{A}$. For this setting, the secrecy-capacity is given by

$$C_s(\sigma_1, \sigma_2, \mathsf{A}) = \max_{P_X : |X| \le \mathsf{A}} I(X; Y_1) - I(X; Y_2) \tag{3}$$

$$= \max_{P_X : |X| \le \mathsf{A}} I(X; Y_1 | Y_2), \tag{4}$$

where $P_X$ denotes the distribution of the input random variable $X$, and where $I(X; Y_i)$ is the mutual information between $X$ and $Y_i$, $i \in \{1, 2\}$. We are interested in studying the input distribution $P_{X^\star}$ that maximizes (4). It can be shown that for $\sigma_1^2 \ge \sigma_2^2$ the secrecy-capacity is equal to zero. Therefore, in the remaining, we assume that $\sigma_1^2 < \sigma_2^2$.

The $P_{X^\star}$ and $C_s(\sigma_1, \sigma_2, \mathsf{A})$, besides a few cases, are in general unknown. The expression of the capacity is an important benchmark in communications theory, and knowing the secrecy-capacity-achieving distribution is important as it is useful for the code or modulation design. Furthermore, the availability of numerical examples of $P_{X^\star}$ may also guide theoretical work by pointing out possible properties of $P_{X^\star}$ that we might want to prove. Therefore, it is both of theoretical and practical need to produce numerical example of $P_{X^\star}$ and $C_s(\sigma_1, \sigma_2, \mathsf{A})$. In this work, we take a numerical approach to computing $P_{X^\star}$ and $C_s(\sigma_1, \sigma_2, \mathsf{A})$ and provide numerical examples of these quantities for various parameter regimes of $(\mathsf{A}, \sigma_1, \sigma_2)$. Data from these simulations are made available at [1].

### A. Notation

Throughout the paper, the deterministic scalar quantities are denoted by lower-case letters and random variables are denoted by uppercase letters.

We denote the distribution of a random variable $X$ by $P_X$. The support set of $P_X$ is denoted and defined as

$$\mathsf{supp}(P_X) = \{ x : \text{for every open set } \mathcal{D} \ni x \text{ we have that } P_X(\mathcal{D}) > 0 \}. \tag{5}$$

The cardinality of $\mathsf{supp}(P_X)$ will be denoted by $|\mathsf{supp}(P_X)|$. The relative entropy between distributions $P$ and $Q$ will be denoted by $\mathsf{D}(P \| Q)$.

Let $g_1$ and $g_2$ be nonnegative functions, then $g_1(x) = O(g_2(x))$ means that there exists a constant $c > 0$ and $x_0$ such that $\frac{g_1(x)}{g_2(x)} \le c$ for all $x > x_0$;

### B. Literature Review

The wiretap channel was introduced by Wyner in [2], who also established the secrecy-capacity of the degraded wiretap channel. The wiretap channel plays a central role in network information theory; the interested reader is referred to [3]–[6] and reference therein for an in-detail treatment of the topic.

The secrecy-capacity of a Gaussian wiretap channel with an average power constraint was shown by Leung and Hellman in [7] where the capacity-achieving input distribution was shown to be Gaussian. The secrecy-capacity of the Gaussian wiretap channel with an amplitude and power constraint was considered by Ozel *et al.* in [8] where the authors showed that the capacity-achieving input distribution is discrete with finitely many support points. Recently, the result of [8] was sharpened in [9] in several ways. First, an explicit upper bound on the number of support points of $P_{X^\star}$ of the following from has been shown:

$$|\mathsf{supp}(P_{X^\star})| \le \rho \frac{\mathsf{A}^2}{\sigma_1^2} + O(\log(\mathsf{A})), \tag{6}$$

where $\rho = (2\mathrm{e}+1)^2 \left(\frac{\sigma_2+\sigma_1}{\sigma_2-\sigma_1}\right)^2 + \left(\frac{\sigma_2+\sigma_1}{\sigma_2-\sigma_1}+1\right)^2$. Second, the following two lower bounds on the cardinality of the support of the following form have been shown:

$$|\mathsf{supp}(P_{X^\star})| \ge \sqrt{1+\frac{\frac{2\mathsf{A}^2}{\pi \mathrm{e}\sigma_1^2}}{1+\frac{\mathsf{A}^2}{\sigma_2^2}}\mathrm{e}^{I(X^\star;Y_2)}} \tag{7}$$

$$\ge \sqrt{1+\frac{\frac{2\mathsf{A}^2}{\pi \mathrm{e}\sigma_1^2}}{1+\frac{\mathsf{A}^2}{\sigma_2^2}}}, \tag{8}$$

where the second bound follows from the trivial bound $\mathrm{e}^{I(X^\star;Y_2)} \ge 1$.

In this work, specifically capitalizing on the upper bound in (6), we propose an algorithm that produces extensive numerical examples for $P_{X^\star}$. A firm upper bound on the size of the support in [9] allows us to move the optimization from the space of probability distributions to the $\mathbb{R}^{2n}$ space, where $n$ is the number of support points of $P_{X^\star}$ (i.e., $n = |\mathsf{supp}(P_{X^\star})|$). Working in $\mathbb{R}^{2n}$ allows us to employ finite-dimensional methods such as the gradient ascent together with a variant of the Blahut-Arimoto algorithm [10] for the wiretap channel [11]. The firm bound $n$ also guarantees that our algorithm will converge to at least a local maximum. To guarantee our solution is close to the global maximum, we also check it against the Karush-Kuhn-Tucker (KKT) conditions.

To the best of our knowledge, there are no existing prior works which focus on generating numerical examples of secrecy-capacity-achieving distribution for the Gaussian wiretap channel in the regime where $P_{X^\star}$ is not available analytically. However, there is extensive prior work on generating examples of the capacity-achieving distribution for the point-to-point channel, which is a special case of the wiretap channel when $\sigma_2 = \infty$; The interested reader is referred to [12]–[14] for examples of such implementations.

### C. Outline and Contributions

The outline and the contribution of the paper are as follows. Section II will present the following: the KKT conditions needed for the optimality of $P_{X^\star}$; the algorithm that will produce numerical examples of $P_{X^\star}$ and other quantities related to $P_{X^\star}$. Section III will present our extensive numerical simulations and the observations surrounding these simulations.

## II. Proposed Algorithm

In this section, we first present the KKT equations, which provide sufficient and necessary conditions for the optimality of $P_{X^\star}$. Second, we present the algorithm which will be used to generate numerical examples of $P_{X^\star}$. The proposed algorithm is the combination of a version of the Blahut-Arimoto algorithm and the gradient ascent algorithm. The key step of the algorithm evaluates the KKT conditions with the candidate solution, which ensures that the solution is close to the true solution.

### A. KKT conditions

The starting point for the design of our algorithm are the following KKT conditions shown in [8].

**Lemma 1.** *The secrecy-capacity-achieving input distribution $P_{X^\star}$ and induced secrecy-capacity-achieving output distributions $P_{Y_1^\star}$ and $P_{Y_2^\star}$ satisfy the following*

$$\Xi(x; P_{X^\star}) = C_s(\sigma_1, \sigma_2, \mathsf{A}), \qquad x \in \mathsf{supp}(P_{X^\star}), \tag{9a}$$

$$\Xi(x; P_{X^\star}) \le C_s(\sigma_1, \sigma_2, \mathsf{A}), \qquad x \in [-\mathsf{A}, \mathsf{A}], \tag{9b}$$

*where for $x \in \mathbb{R}$*

$$\Xi(x; P_{X^\star}) = \mathsf{D}(P_{Y_1|X}(\cdot|x) \| P_{Y_1^\star}) - \mathsf{D}(P_{Y_2|X}(\cdot|x) \| P_{Y_2^\star}). \tag{10}$$

In the numerical computation of $P_{X^\star}$ we find more convenient to check the negated version of (9), where a tolerance parameter $\varepsilon$ is introduced which trades off accuracy with computational burden. Specifically, $P_X$ is not an optimal input pmf if any of the following conditions is satisfied:

$$\Xi(\mathsf{A}; P_X) + \varepsilon < \Xi(x; P_X), \text{ for some } x \in [-\mathsf{A}, \mathsf{A}] \tag{11a}$$

$$|\Xi(x; P_X) - \Xi(\mathsf{A}; P_X)| > \varepsilon, \text{ for some } x \in \mathsf{supp}(P_X). \tag{11b}$$

Note that in (11) in place of the secrecy-capacity $C_s(\sigma_1, \sigma_2, \mathsf{A})$, which is unknown, we used the value of $\Xi(\mathsf{A}; P_X)$, thanks to the fact that $\mathsf{A} \in \mathsf{supp}(P_{X^\star})$ for any $(\sigma_1, \sigma_2, \mathsf{A})$. With some abuse of notation, we refer to (11) as to the $\varepsilon$-KKT conditions.

### B. Numerical Algorithm

The algorithm that gives a numerical lower bound to secrecy-capacity and that estimates the optimal input pmf $P_{X^\star}$ is given in Algorithm 1. The algorithm takes as input: the value A of the amplitude constraint; the values $(\sigma_1, \sigma_2)$ of the standard deviations of the additive noise; the vector $\mathbf{x}$ of the support points of the initial tentative $P_X$; the vector $\mathbf{p}$ of the probabilities associated with the points in $\mathbf{x}$; and a value $\varepsilon$ which is related to the accuracy of the capacity estimation. Using the bound in [9, Eq. (83)], we can set the dimensionally of $\mathbf{p}$ and $\mathbf{x}$ to be less than

$$b_1 \frac{\mathsf{A}^2}{\sigma_1^2} + b_2 + \log \frac{b_3\mathsf{A} + b_4\mathsf{A} + b_5}{b_6\mathsf{A} + b_7}, \tag{12}$$

where the constants $\{b_i\}_{i=1}^7$ depend on $(\sigma_1, \sigma_2, C_s)$. We can easily prove that if $x \in \mathsf{supp}(P_{X^\star})$, then $-x \in \mathsf{supp}(P_{X^\star})$ and

**Algorithm 1** Capacity and input PMF estimation

**procedure** $\textsc{Main}(\mathsf{A}, \sigma_1, \sigma_2, \mathbf{x}, \mathbf{p}, \varepsilon)$
  $N_{\text{BA}} \leftarrow 100$ / *Number of Blahut-Arimoto iterations*
  $N_{\text{GA}} \leftarrow 20$ / *Number of gradient ascent iterations*
  **repeat**
    $k \leftarrow 0$
    **while** $k < 100$ **do**
      $k \leftarrow k + 1$
      $\mathbf{p} \leftarrow \textsc{Blahut-Arimoto}(\mathbf{x}, \mathbf{p}, N_{\text{BA}})$
      $\mathbf{x} \leftarrow \textsc{Gradient-Ascent}(\mathbf{x}, \mathbf{p}, N_{\text{GA}})$
    **end while**
    $(\mathbf{x}, \mathbf{p}) \leftarrow \textsc{Cluster}(\mathbf{x}, \mathbf{p})$
    valid $\leftarrow \textsc{KKT-Validation}(\mathbf{x}, \mathbf{p}, \varepsilon)$
    **if** valid = False **then**
      $(\mathbf{x}, \mathbf{p}) \leftarrow \textsc{Update}(\mathbf{x}, \mathbf{p})$
    **end if**
  **until** valid = True
  $C_s(\sigma_1, \sigma_2, \mathsf{A}) \leftarrow \Xi(\mathbf{x}, \mathbf{p})$
  **return** $\mathbf{x}, \mathbf{p}, C_s(\sigma_1, \sigma_2, \mathsf{A})$
**end procedure**

$P_{X^\star}(x) = P_{X^\star}(-x)$; We can exploit the symmetry by storing in $\mathbf{x}$ only the nonnegative valued support points.

The optimization procedure is iterative and divided into two parts: (i) $N_{\text{BA}}$ iterations of the Blahut-Arimoto algorithm are used to let the tentative pmf converge to stable values, and (ii) $N_{\text{GA}}$ iterations of a gradient ascent algorithm are used to modify the positions of the support points of the tentative input distribution. The objective function of the gradient ascent is the secrecy-information $I(X;Y_1) - I(X;Y_2)$. We use a backtracking line search version of the gradient ascent [15], in order to ensure that the sequence of secrecy-information values obtained along the iterations is nondecreasing, which ensures convergence to a local maximum.

After the two-step optimization, if there is a set of support points that are driven too close to each other by the gradient ascent, then such points are clustered together in a new support point which assumes a probability equal to the sum of the probabilities of the clustered points. Note that the clustering operation is equivalent to having a minimum distance constraint among support points. We have observed that this constraint in general helps numerical stability. We have chosen a minimum distance of $10^{-2}$, which seems to be an inactive capacity constraint for all values of $\mathsf{A} > 10^{-2}$ and of $(\sigma_1, \sigma_2)$. We caution, however, this constraint might be active around the transition points (i.e., when one point splits into two), but becomes inactive once again after we move away by $10^{-2}$ from the transition point.

The optimality of the resulting distribution $P_X$ is tested with the $\varepsilon$-KKT conditions. If any of the conditions (11) is satisfied, then the function $\textsc{Update}$ modifies $P_X$ according to the following rules. Let

$$\widehat{x} = \arg \max_{x \in [-\mathsf{A}, \mathsf{A}]} \Xi(x; P_X) \tag{13}$$

be the candidate novel support point (or points, due to the even symmetry of $\Xi(\cdot; P_X)$), and

$$\mathcal{S} = \{x \in \mathsf{supp}(P_X) : \text{(11b) is true}\} \tag{14}$$

be the set of support points for which $\Xi(\cdot; P_X)$ falls outside the horizontal $\varepsilon$-strip that contains $\Xi(\mathsf{A}; P_X)$. If both conditions in (11) are verified and there exist $x_1, x_2 \in \mathcal{S}$ such that $|x_1 - x_2| < \delta = 0.1$ and $\widehat{x} \in [x_1, x_2]$, then $\widehat{x}$ replaces both $x_1$ and $x_2$ in the support of $P_X$ with $P_X(\widehat{x}) = P_X(x_1) + P_X(x_2)$. Otherwise, if only (11a) is satisfied, then $\widehat{x}$ is added to the support of $P_X$ and the probabilities are set to $P_X(x) = |\mathsf{supp}(P_X)|^{-1}$ for all $x \in \mathsf{supp}(P_X)$.

The main algorithm repeats the two-step optimization procedure until $P_X$ is validated, which becomes the proposed $P_{X^\star}$.

## III. Simulations and Observations

In this section, we use Algorithm 1 to provide examples of the $P_{X^\star}$. For several values of the parameters $(\sigma_1, \sigma_2, \mathsf{A})$, we will plot the following: $C_s(\sigma_1, \sigma_2, \mathsf{A})$; the location of the support points of $P_{X^\star}$; the cardinality of $\mathsf{supp}(P_{X^\star})$; the output pdfs $P_{Y_1^\star}$ and $P_{Y_2^\star}$. The simulation data results are publicly available [1].

We consider two regimes. In the first regime we examine a scenario where the eavesdropper experiences significantly large noise than the legitimate user and specifically we consider $\sigma_1^2 = 1$ and $\sigma_2^2 = 10$. In the second regime, we examine a scenario where the noise at both users is comparable and specifically we consider $\sigma_1^2 = 1$ and $\sigma_2^2 = 1.5$.

Fig. 2 presents the structure of the support of the secrecy-capacity-achieving distribution vs. $\mathsf{A}$ and the values of $C_s(\sigma_1, \sigma_2, \mathsf{A})$ vs. $\mathsf{A}$. We observe the following:

- In Fig. 2a we plot $\mathsf{supp}(P_{X^\star})$ vs. $\mathsf{A}$, where due to symmetry of $P_{X^\star}$ we only plot the nonnegative points in $\mathsf{supp}(P_{X^\star})$. In addition, in Fig. 2b we normalize the values of the support and plot the nonnegative values of $\frac{\mathsf{supp}(P_{X^\star})}{\mathsf{A}}$ vs. $\mathsf{A}$. First, we see that, as $\mathsf{A}$ increases, the new points appear only at zero while the nonzero points travel away from the origin. Moreover, the points are created in only the following two ways: either a point at zero simply emerges, or an existing point at zero splits into two points. This behavior is analogous to the behavior of the support of the capacity-achieving distribution for the point-to-point channel [16].
  Second, note that the support points of $P_{X^\star}$ for $\sigma_2^2 = 10$ leg behind (*i.e.*, for larger values of $\mathsf{A}$) the support points of $P_{X^\star}$ for $\sigma_2^2 = 1.5$. This could be explained by noting that when the eavesdropper faces less additive noise, the transmitter has to create more equivocation in order to keep the leakage information rate at the eavesdropper equal to zero: This could be done only by adding a new support point in $P_{X^\star}$.
- In Fig. 2c we plot the cardinality of $\mathsf{supp}(P_{X^\star})$ vs. $\mathsf{A}$. First, we observe that the cardinality increases linearly in $\mathsf{A}$. Should this behavior continue for larger values of $\mathsf{A}$, the upper bound in (6) will be loose. Note a similar issue exists for a point-to-point channel with a peak amplitude constraint where the cardinality of the support of the capacity-achieving distribution is lower-bounded by a term in the order of A while the upper bound is in the order of $\mathsf{A}^2$ [16]. Therefore, since the point-to-point channel is a special case of the wiretap channel, it is not surprising that similar behavior occurs here too.

Second, Fig. 2c suggests that the bound in (7) might be loose. However, we conjecture that this bound, while it may not be attaining the correct constants, is tight asymptotically as A increases.

- In Fig. 2d we plot the difference between the nonnegative adjacent points in $\mathsf{supp}(P_{X^\star})$ starting from the largest and the second largest. An interesting observation here is that the distance between two adjacent points eventually stops fluctuating and concentrates on a single value. In other words, the space between adjacent points eventually does not change. However, it is important to note that the spacing is not uniform. Along this direction, an interesting future direction would be to consider the asymptotic behavior of $\frac{X^\star}{\mathsf{A}}$ and find the limiting distribution as $\mathsf{A} \to \infty$.
- In Fig. 2e we plot the secrecy-capacity vs. A. It is interesting to note that the secrecy-capacity is not concave for small values A. There are two reasons for this. First, note that $C_s$ is given as the difference of mutual informations. Therefore, even if each mutual information is concave in A, this does not imply that $C_s$ is concave in A as the difference of concave functions is not concave. Second, note that even the single mutual information may not be concave in A: See Fig. 2f where the mutual information is plotted vs. A. Note that because of the I-MMSE relationship [17], the mutual information is a concave function of the signal-to-noise ratio and is, therefore, concave in $\mathsf{A}^2$. However, this does not imply that the mutual information is concave in A. In fact, as we see from Fig. 2f, for small values of A, the mutual information can be convex in A.

In addition to studying the optimal input distribution, it is also interesting to observe the behavior of the output distributions $P_{Y_1^\star}$ at the legitimate user and $P_{Y_2^\star}$ at the eavesdropper, and to compare them. Fig. 3 provides examples of $P_{Y_1^\star}$, $P_{Y_2^\star}$, and $P_{X^\star}$ for $\sigma_1^2 = 1$ and $\sigma_2^2 = 10$, and Fig. 4 provides examples of $P_{Y_1^\star}$, $P_{Y_2^\star}$, and $P_{X^\star}$ for $\sigma_1^2 = 1$ and $\sigma_2^2 = 1.5$.

The eavesdropper's pdf $P_{Y_2^\star}$ and the legitimate user's pdf $P_{Y_1^\star}$ have clear differences. In particular, $P_{Y_2^\star}$ has just one inflection point for positive abscissa values, and very much resembles the Gaussian pdf. To aid this comparison of $P_{Y_2^\star}$ to the Gaussian distribution, we also plot $P_{Y_G}$ where $Y_G$ is the Gaussian random variable with variance $\mathbb{E}[(X^\star)^2] + \sigma_2^2$ (*i.e.*, $Y_G$ and $Y_2^\star$ have the same variance). In contrast, $P_{Y_1^\star}$ has many peaks and inflection points, which correspond to the locations of the support points of $P_{X^\star}$. Because of the difference in structure of the two pdfs, estimating $X^\star$ is *easier* from the sample $Y_1^\star$ than from the sample $Y_2^\star$. Of course, this is expected, as we desire the eavesdropper to gain little information from the observation $Y_2^\star$. The proximity to Gaussian can also be seen through the mutual information. In particular, Fig. 2f plots $I(X^\star; Y_2^\star)$ and compares it to the mutual information attained by the Gaussian input with same variance as $X^\star$.

As one possible future direction, it would be interesting to show that $P_{Y_2^\star}$ is indeed close to Gaussian in some distance between distributions (e.g., relative entropy). It would also be interesting to study how the smoothnesses of $P_{Y_1^\star}$ and $P_{Y_2^\star}$ differ. This can be accomplished by studying the maximum number of inflection points of each pdf.

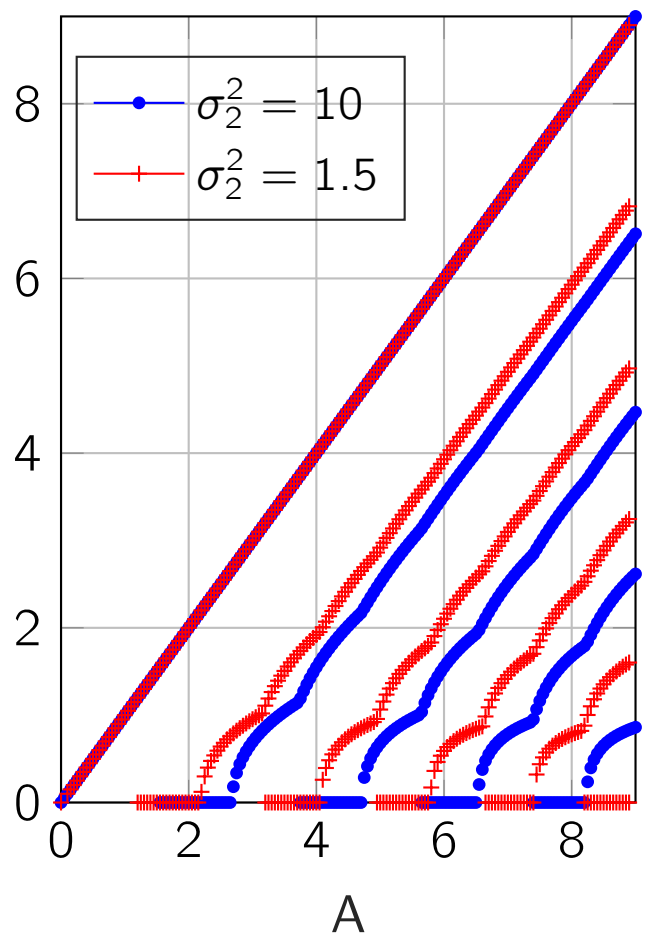

(a) Plot of the nonnegative values of $\mathbf{supp}(P_{X^\star})$ vs. A.

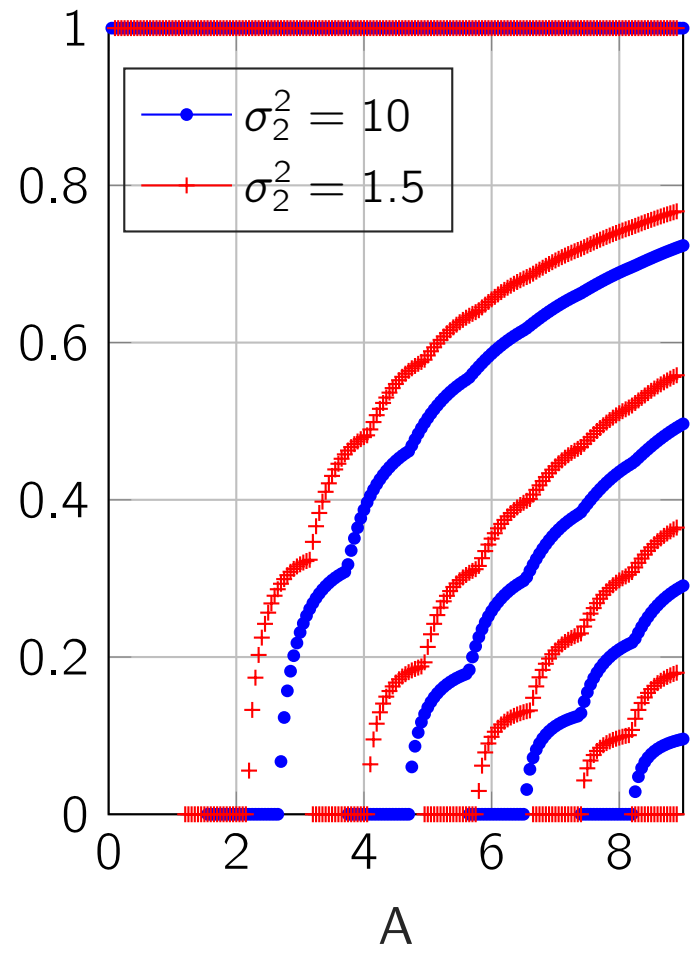

(b) Plot of the nonnegative values of $\frac{\mathbf{supp}(P_{X^\star})}{\mathsf{A}}$ vs. A.

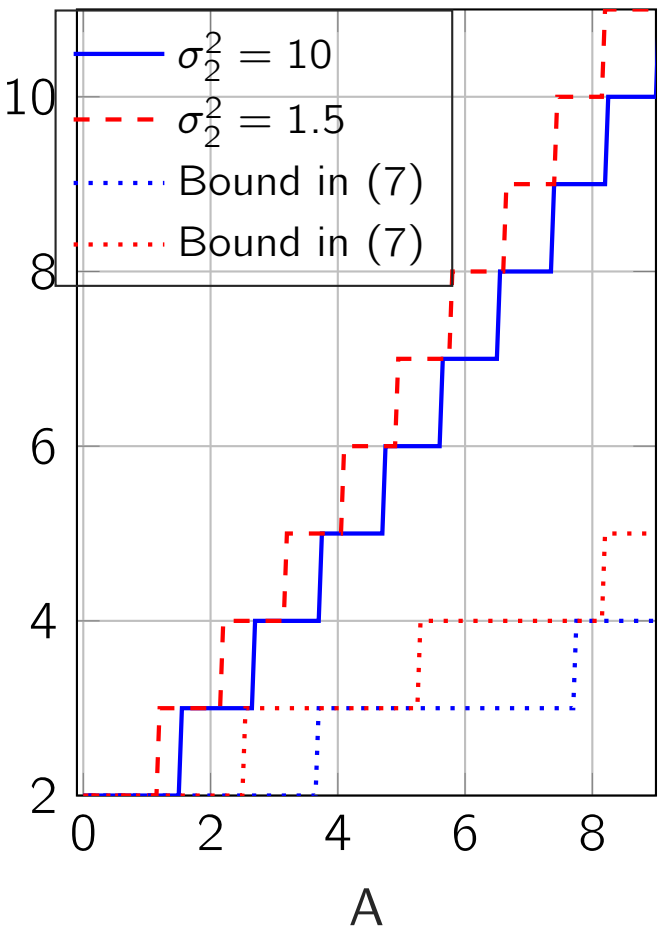

(c) Plot of $|\mathbf{supp}(P_{X^\star})|$ vs. A.

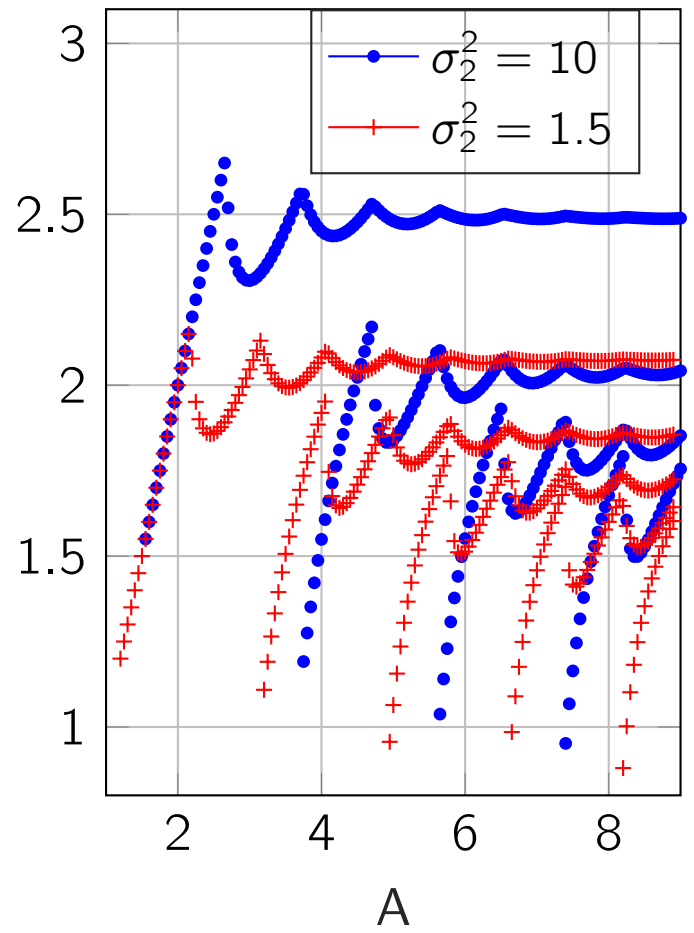

(d) Distance between adjacent points of $\mathbf{supp}(P_{X^\star})$ vs. A.

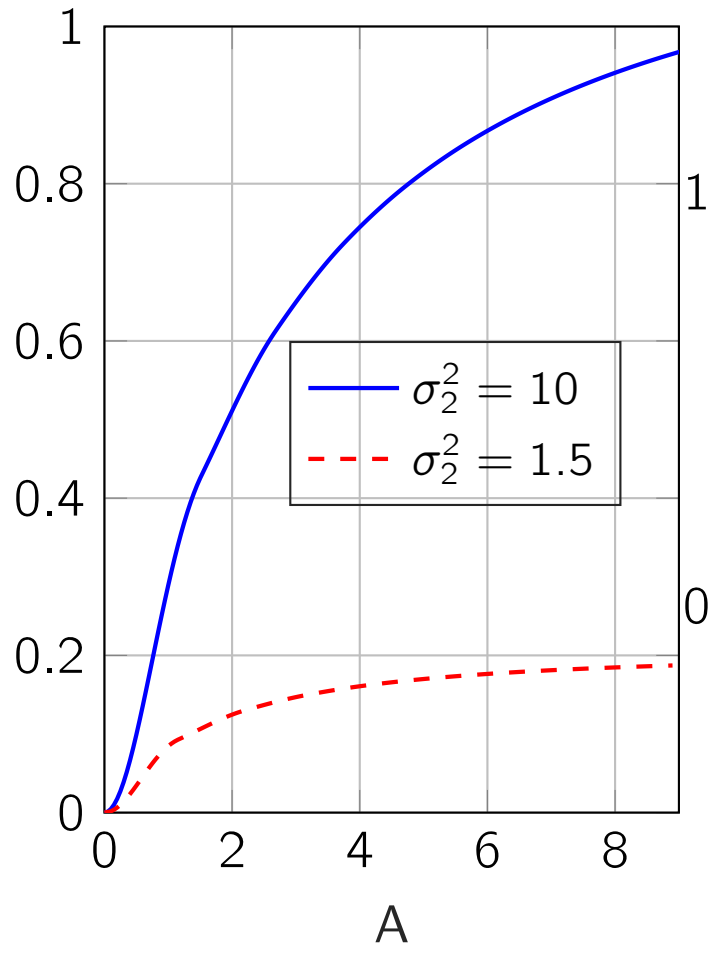

(e) Plot of $C_s(\sigma_1, \sigma_2, \mathsf{A})$ (in nats) vs. A.

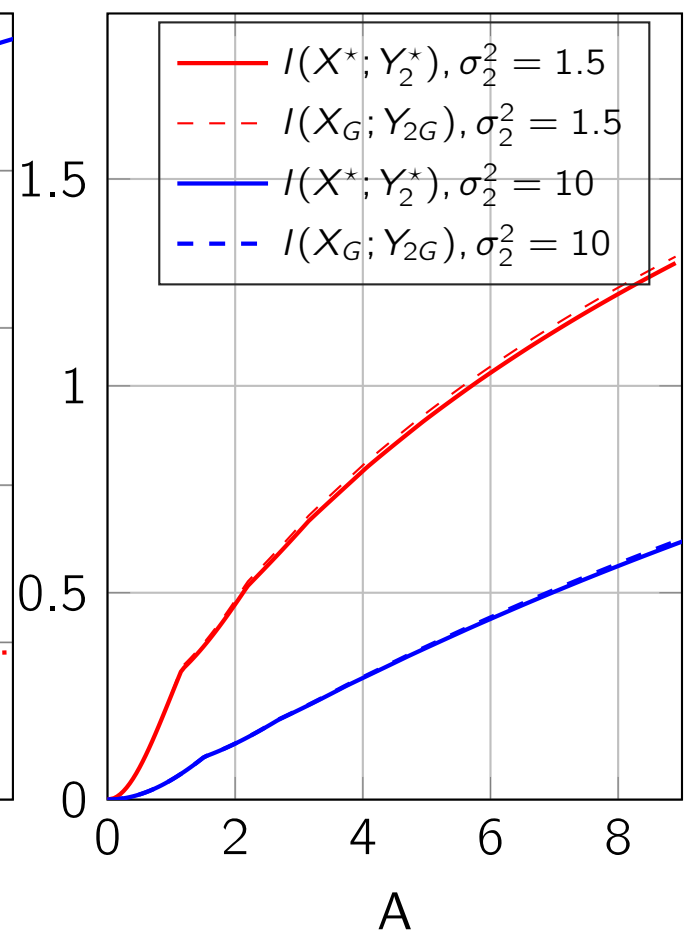

(f) Plot of $I(X^\star; Y_2^\star)$ vs. A.

Fig. 2: Simulations results for the case of $\sigma_1^2 = 1$ and $\sigma_2^2 = \{1.5, 10\}$.

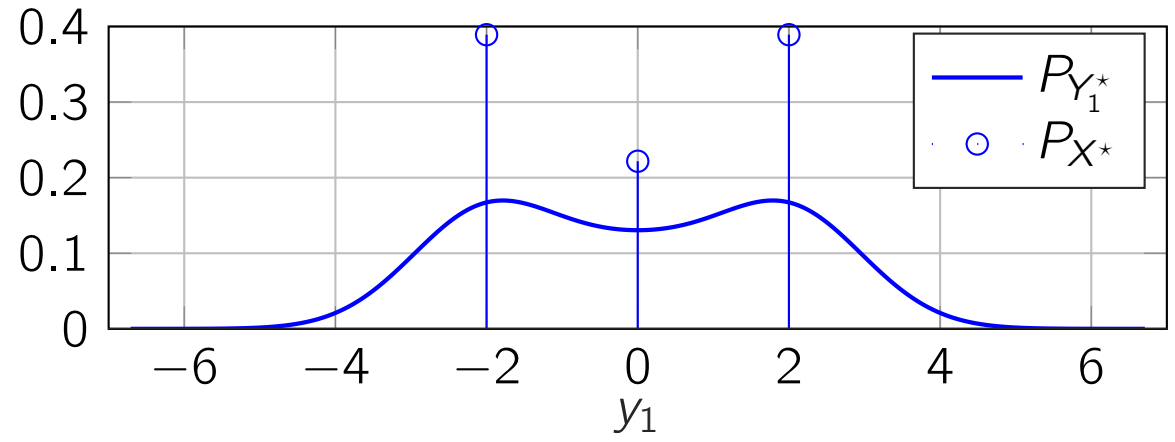

(a) Plot of $P_{Y_1^\star}$ (pdf at legitimate user) and $P_{X^\star}$ for A = 2.

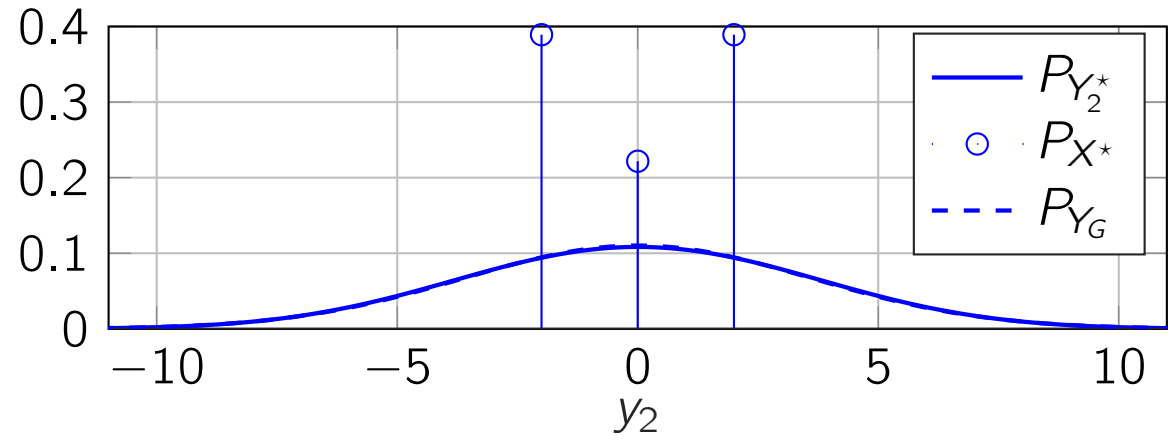

(b) Plot of $P_{Y_2^\star}$ (pdf at eavesdropper) and $P_{X^\star}$ for A = 2.

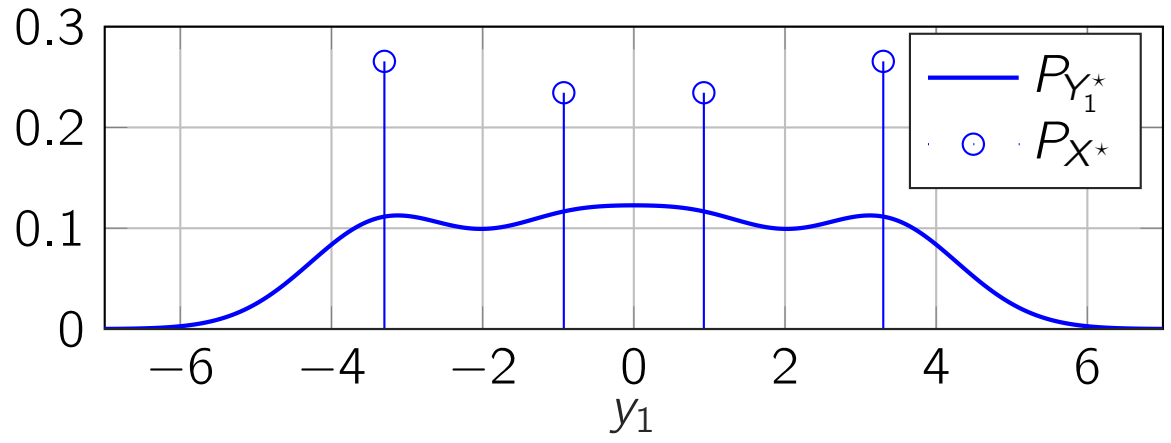

(c) Plot of $P_{Y_1^\star}$ (pdf at legitimate user) and $P_{X^\star}$ for A = 3.3.

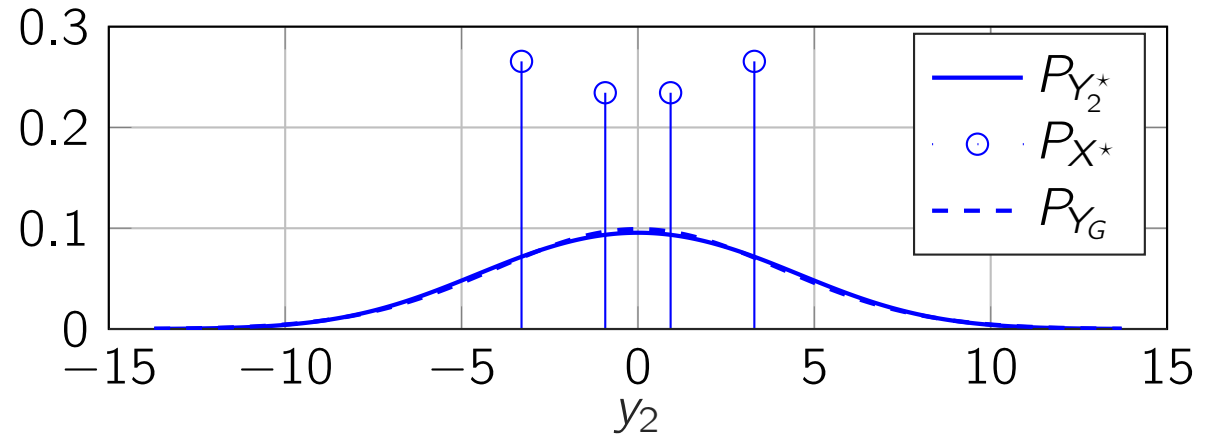

(d) Plot of $P_{Y_2^\star}$ (pdf at eavesdropper) and $P_{X^\star}$ for A = 3.3.

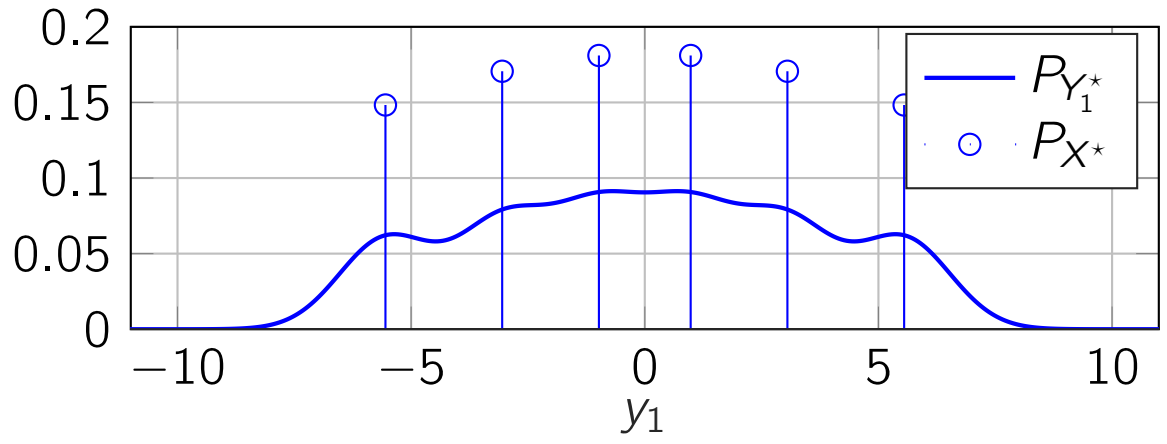

(e) Plot of $P_{Y_1^\star}$ (pdf at legitimate user) and $P_{X^\star}$ for A = 5.55.

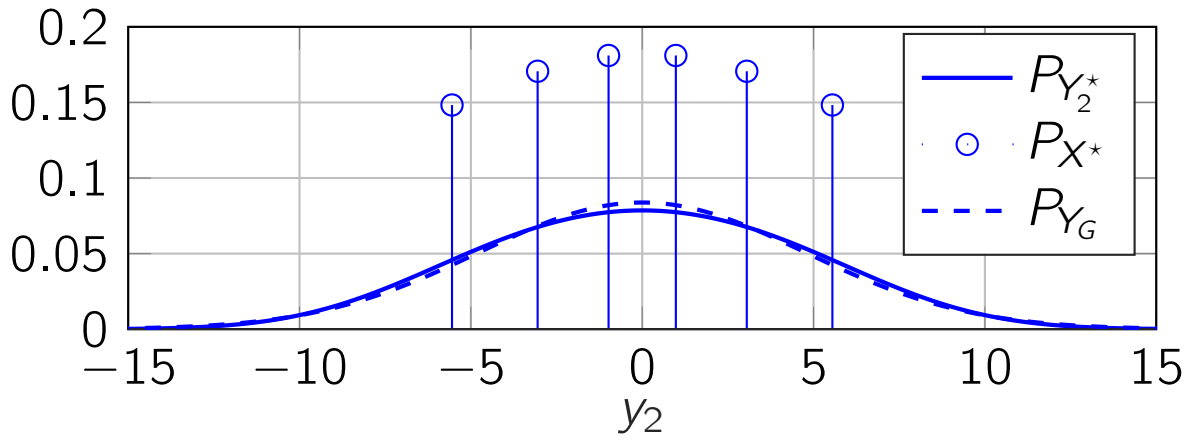

(f) Plot of $P_{Y_2^\star}$ (pdf at eavesdropper) and $P_{X^\star}$ for A = 5.55.

Fig. 3: Examples of input and output distributions for the case of $\sigma_1^2 = 1$ and $\sigma_2^2 = 10$.

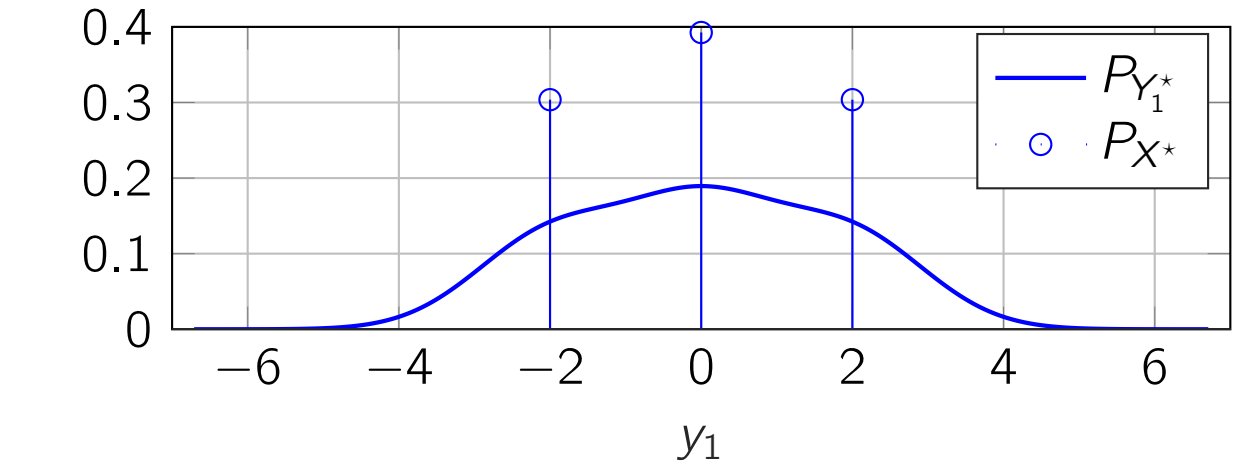

(a) Plot of $P_{Y_1^\star}$ (pdf at legitimate user) and $P_{X^\star}$ for A $= 2$.

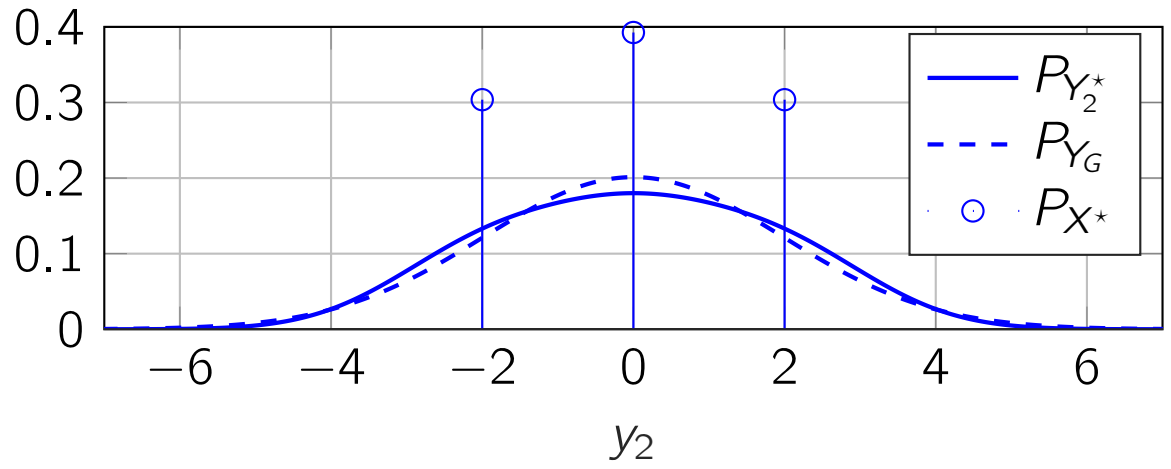

(b) Plot of $P_{Y_2^\star}$ (pdf at eavesdropper) and $P_{X^\star}$ for A $= 2$.

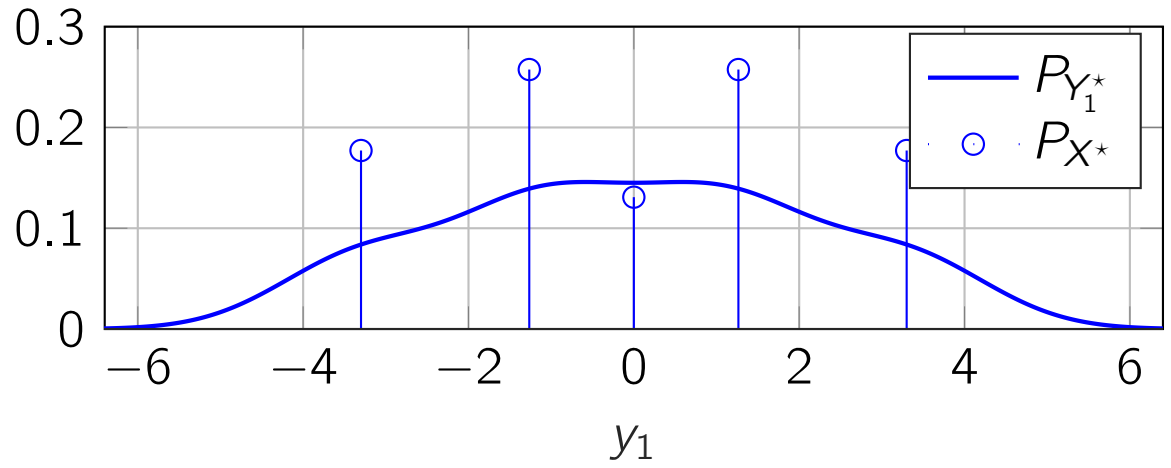

(c) Plot of $P_{Y_1^\star}$ (pdf at legitimate user) and $P_{X^\star}$ for A $= 3.3$.

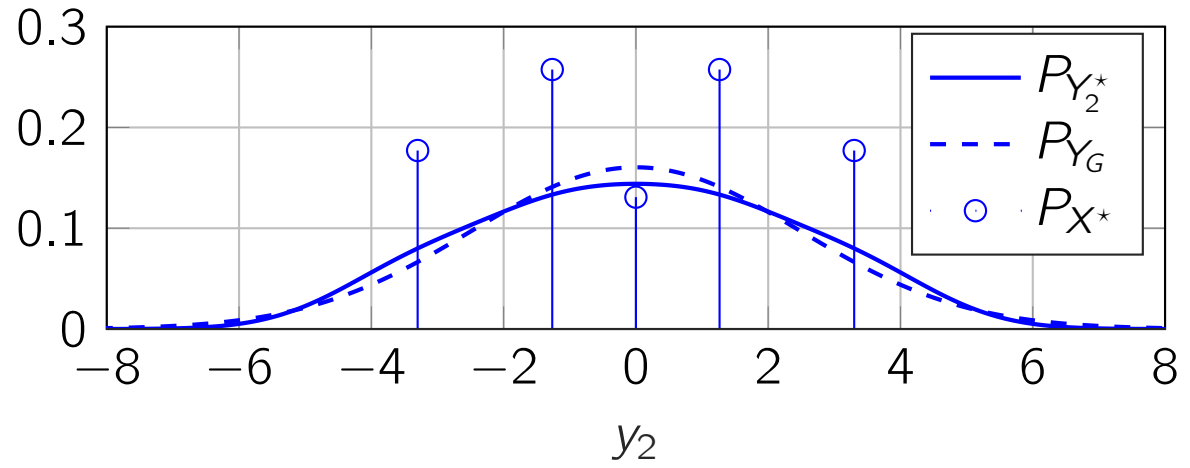

(d) Plot of $P_{Y_2^\star}$ (pdf at eavesdropper) and $P_{X^\star}$ for A $= 3.3$.

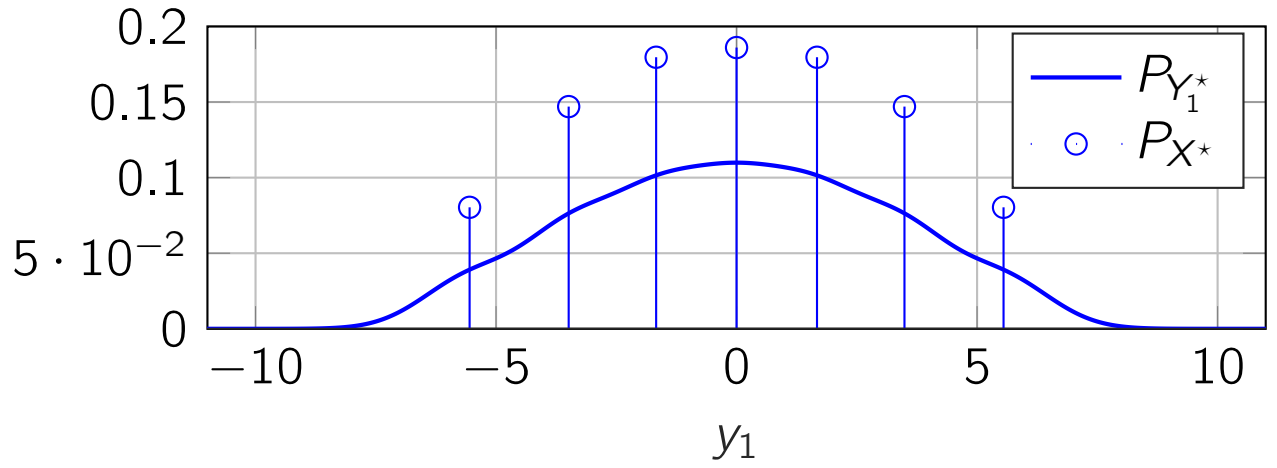

(e) Plot of $P_{Y_1^\star}$ (pdf at legitimate user) and $P_{X^\star}$ for A $= 5.55$.

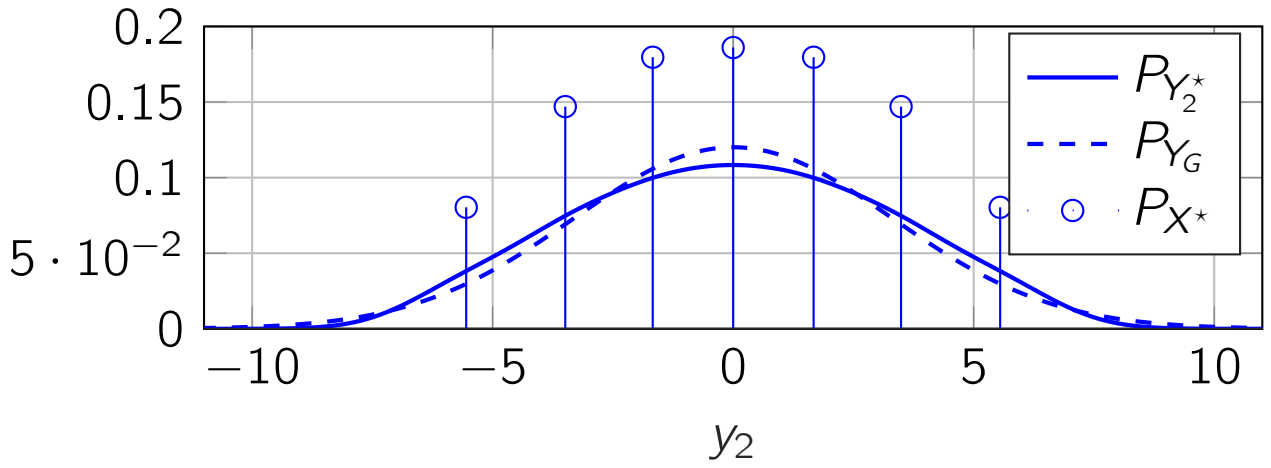

(f) Plot of $P_{Y_2^\star}$ (pdf at eavesdropper) and $P_{X^\star}$ for A $= 5.55$.

Fig. 4: Examples of input and output distributions for the case of $\sigma_1^2 = 1$ and $\sigma_2^2 = 1.5$.